\documentclass[twocolumn,prc,superscriptaddress,unsortedaddress,preprintnumbers,amsmath,amssymb]{revtex4}
\usepackage{tipa}

\usepackage{blindtext}
\usepackage{graphicx}
\usepackage{color}
\usepackage{times}
\usepackage{bm}
\usepackage{multirow}
\usepackage{float}
\usepackage{url}
\usepackage{natbib}
\usepackage{tensor}
\usepackage{bbm}
\usepackage[normalem]{ulem}
\usepackage{verbatim}
\usepackage{lipsum} 
\usepackage{tikz,xcolor}
\usepackage{ifxetex,ifluatex}
\usepackage{booktabs}
\usepackage[colorlinks=true,citecolor=blue,urlcolor=blue,linkcolor=red]{hyperref}

\usepackage[utf8]{inputenc}
\usepackage{dcolumn}% Align table columns on decimal point
\def\bal#1\eal{\begin{align}#1\end{align}}

\def\beq{\begin{equation}}
\def\eeq{\end{equation}}
\def\bea{\begin{eqnarray}}
\def\eea{\end{eqnarray}}

\def\fun#1#2{\lower3.6pt\vbox{\baselineskip0pt\lineskip.9pt
  \ialign{$\mathsurround=0pt#1\hfil##\hfil$\crcr#2\crcr\sim\crcr}}}
\usepackage[normalem]{ulem}  %  \sout 

\newcommand{\physrep}{\textit{Phys.~Rep.}}
\newcommand{\nphysa}{\textit{Nucl.~Phys.~A}}

\preprint{}

\begin{document}

\title{In-medium effects of nucleon-nucleon cross sections in heavy-ion collisions}
\author{Shuochong Han}
\affiliation{Department of Astronomy, Xiamen University, Xiamen, Fujian 361005, China}
\affiliation{Institute of Modern Physics, Chinese Academy of
Sciences, Lanzhou 730000, China}

\author{Xinle Shang}\email[ ]{shangxinle@impcas.ac.cn}
\affiliation{Institute of Modern Physics, Chinese Academy of
Sciences, Lanzhou 730000, China} \affiliation{School of Nuclear
Science and Technology, University of Chinese Academy of Sciences,
Beijing 100049, China}

\author{Wei Zuo}
\affiliation{Institute of Modern Physics, Chinese Academy of
Sciences, Lanzhou 730000, China} \affiliation{School of Nuclear
Science and Technology, University of Chinese Academy of Sciences,
Beijing 100049, China}

\author{Gaochan Yong}
\affiliation{Institute of Modern Physics, Chinese Academy of
Sciences, Lanzhou 730000, China} \affiliation{School of Nuclear
Science and Technology, University of Chinese Academy of Sciences,
Beijing 100049, China}

\author{Ang Li}
\affiliation{Department of Astronomy, Xiamen University, Xiamen, Fujian 361005, China}

\begin{abstract}

Based on the isospin-dependent Boltzmann-Uehling-Uhlenbeck transport model, we systematically investigate the in-medium effects of nucleon–nucleon ($NN$) cross sections on nucleonic and pionic observables in heavy-ion collisions, employing microscopic cross sections derived from the Brueckner–Hartree–Fock approach. Key observables include nuclear stopping, the neutron-to-proton ($n/p$) ratio, neutron–proton transverse flow differences, differential collective flow, pion multiplicities, and the resulting $(\pi^-/\pi^+)_{\text{like}}$ ratio. The analysis disentangles the respective contributions from the scattering amplitude, the density of states, and the total momentum ($K$) of the colliding pairs. We find that larger in-medium $NN$ cross sections generally enhance free nucleon emission and nuclear stopping, with the nucleon effective mass playing a dominant suppressive role. However, it is insufficient to account only for the medium corrections from effective mass: both the medium effect from the scattering amplitude and the $K$-dependence exert noticeable influences on the observables. In particular, nuclear stopping is found to be highly sensitive to these in-medium modifications of cross sections.
While the $n/p$ ratio and transverse flow difference remain largely insensitive, the differential collective flow and pion yields are strongly affected. These results indicate that the interplay between scattering amplitude, density-of-states and $K$-dependence is essential to accurately describe medium effects in heavy-ion collisions.

\end{abstract}

\maketitle

\section{INTRODUCTION}
The in-medium nucleon-nucleon ($NN$) cross section constitutes a critical input for transport models used to simulate heavy-ion collisions (HICs) at intermediate energies \cite{1993PhRvL..71.1986W,1995PhRvC..51.3320S,1995NuPhA.587..815A,1994PhLB..334...12I,1992PhRvC..46..677D,2010EPJA...46..399Y,2022PhLB..82837019L}. These collisions serve as a key probe of the high-density behavior of the nuclear equation of state, providing valuable insights into the properties of nuclear matter under extreme conditions \cite{2018PhLB..778..207W,2008PhR...464..113L,2024PhRvC.109e4619L}. Observables that are particularly sensitive to the equation of state, such as nucleon collective flow \cite{2000PhRvL..85.4221L,2006PhRvC..74f4617Y,2014PhRvC..89d4603W} and pion production \cite{2005PhRvC..71a4608L,2006PhRvC..73c4603Y,2009PhRvL.102f2502X}, may also exhibit strong dependence on the in-medium $NN$ cross section. Furthermore, transport models, including Boltzmann-Uehling-Uhlenbeck (BUU) and quantum molecular dynamics (QMD), incorporate two essential ingredients: the nuclear mean field \cite{2022PrPNP.12503962W,2021PhRvC.104b4603C} and the in-medium $NN$ cross section \cite{2016PhRvC..93d4609X,2018PhRvC..97c4625Z}. The mean field describes the average single-particle potential experienced by nucleons, whereas the in-medium cross section accounts for the medium modifications of $NN$ interactions. A self-consistent description of HIC dynamics requires both ingredients to be derived from microscopic many-body theories based on realistic $NN$ interactions.

However, unlike the $NN$ cross section in free space $\sigma_{\text{free}}$, which is well constrained by scattering experiments, its in-medium counterpart cannot be directly measured due to the complex dynamics and correlations among nucleons in dense nuclear matter. Consequently, theoretical predictions based on $ab\ initio$ calculations are indispensable for constraining the in-medium $NN$ cross section. Microscopic approaches such as the Brueckner theory \cite{1989NuPhA.494..349B,1997PhRvC..55.3006S,2021PhRvC.103c4316S}, Dirac-Brueckner theory \cite{1993PhRvC..48.1702L,1994PhRvC..49..566L,2001PhRvC..64b4003F,2006PhRvC..73a4001S}, and variational methods \cite{1992PhRvC..45..791P}, have been extensively employed for this purpose. Among these, the Brueckner–Hartree–Fock (BHF) framework provides a systematic approach to calculating in-medium $NN$ cross sections by incorporating two primary medium effects: modifications to the scattering amplitude and changes in the density of states.
The scattering amplitude is represented by the effective $G$-matrix, which encapsulates medium corrections such as Pauli blocking and mean-field potentials, while the density of states is influenced by the effective masses of the nucleons \cite{1997PhRvC..55.3006S}. Importantly, the in-medium $NN$ cross section depends not only on the relative momentum $k$ and the scattering angle $\theta$ of the colliding nucleons, but also on their total momentum $K$. This $K$-dependence originates from an accurate treatment of the Brueckner-Bethe-Goldstone (BBG) equation \cite{2021PhRvC.103c4316S},  
highlighting the sensitivity of in-medium scattering processes to the motion of the nucleon pair relative to the medium. Moreover, the $K$-dependence plays a crucial role in predicting the isospin dependence of the nucleon effective mass \cite{2020PhRvC.101f5801S}, which in turn significantly influences the in-medium $NN$ cross section. In our previous work \cite{2022PhRvC.106f4332H,2021PhRvC.103c4316S}, we presented detailed microscopic predictions of in-medium $NN$ cross sections in asymmetric nuclear matter within the BHF framework, fully accounting the dependence on the total momentum $K$ of the colliding nucleon pair-i.e., incorporating its effects in both the density of state and the scattering amplitude.

To integrate these microscopic results into practical transport simulations, a commonly adopted approximation in isospin-dependent Boltzmann–Uehling–Uhlenbeck (IBUU) transport model is to factorize the in-medium $NN$ cross section as the product of the free-space cross section and a medium correction factor. 
In previous studies\cite{2005PhRvC..72f4611L,2013PhLB..726..211G,2021PhRvC.103c4615Z}, this factor is typically expressed as the square of the ratio between the in-medium effective reduced mass and the bare reduced mass of the colliding pair, providing only an approximate treatment of the medium effects by accounting for changes in the density of states while neglecting the influence of medium modifications to the scattering amplitude. 
In contrast, in Ref. \cite{2022PhRvC.106f4332H}, we proposed analytical expressions for the medium correction factors based on microscopic BHF calculations, explicitly incorporating the modifications on both the density of states and the scattering amplitude. 
However, no systematic study has been carried out to evaluate the impact of in-medium $NN$ cross sections with accurately treated medium effects on the HIC observables. To address this, here we extend our previously proposed correction factors to systematically investigate the in-medium effects of $NN$ cross sections on several key observables sensitive to the cross section in intermediate-energy HICs, including the nuclear stopping, the neutron-to-proton $(n/p)$ ratio of free nucleons, pion production, and neutron-proton differential transverse and elliptic flows. Particular emphasis is placed on disentangling the contributions from medium modifications to the scattering amplitude, to the density of states, and to the $K$-dependence. Our results provide further insight into the importance of consistent microscopic treatment of in-medium $NN$ cross sections in transport model simulations.

The paper is organized as follows. In the next section, we briefly review the IBUU method and the in-medium $NN$ cross sections based on BHF calculations. 
Sec. III consists of three parts: the first subsection focuses on the numerical results and interpretation of nuclear stopping; the second discusses the neutron-proton differential flow obtained using microscopic in-medium cross sections; and the third analyzes the production yields of $\pi^-$ and $\pi^+$ mesons. 
And finally, a summary is given in Sec. IV.

\section{THEORETICAL APPROACHES}

In this study, we utilize the semiclassical transport model known as the IBUU model, where the scattering between nucleons is simulated through the collision integral, whose expression is 
\begin{eqnarray}
    I_c&=&\frac{4}{(2\pi)^3}\int\int d\emph{\textbf{p}}_{2}d\emph{\textbf{p}}_{3}\int d\Omega|\emph{\textbf{v}}_{12}|\frac{d\sigma_{NN}}{d\Omega}(\emph{\textbf{p}}_2-\emph{\textbf{p}}_4)\nonumber\\
    &\times&\delta(\emph{\textbf{p}}_1+\emph{\textbf{p}}_2-\emph{\textbf{p}}_3-\emph{\textbf{p}}_4)\{f(\emph{\textbf{r}},\emph{\textbf{p}}_3,t)f(\emph{\textbf{r}},\emph{\textbf{p}}_4,t)\nonumber\\
    &\times&[1-f(\emph{\textbf{r}},\emph{\textbf{p}}_1,t)][1-f(\emph{\textbf{r}},\emph{\textbf{p}}_2,t)]-f(\emph{\textbf{r}},\emph{\textbf{p}}_1,t)\nonumber\\
    &\times&f(\emph{\textbf{r}},\emph{\textbf{p}}_2,t)[1-f(\emph{\textbf{r}},\emph{\textbf{p}}_3,t)][1-f(\emph{\textbf{r}},\emph{\textbf{p}}_4,t)]\} \ , 
\end{eqnarray}
where $f(\emph{\textbf{r}},\emph{\textbf{p}},t)$ is phase space distribution function, $[1-f(\emph{\textbf{r}},\emph{\textbf{p}}_{1/3},t)][1-f(\emph{\textbf{r}},\emph{\textbf{p}}_{2/4},t)]$ represents the quantum effect of Pauli blocking. $\sigma_{NN}$ is the particle scattering cross section, including the $NN$ elastic scattering cross section. 
$\emph{\textbf{v}}_{12}$ denotes the relative velocity of the colliding pair and it is independent of the nucleon effective mass in the current work.
Although this treatment is widely used in BUU-type simulations, it represents a simplification and should be acknowledged as a limitation of the present study.
In the IBUU transport model, the effective $NN$ elastic scattering cross section modified by effective mass, is often expressed as follows \cite{2005PhRvC..72f4611L,2014PhRvC..90d4605G}
\begin{eqnarray}
    \sigma_{\text{eff}}=R*\sigma_{\text{free}}=(\frac{\mu^*}{\mu})^2\sigma_{\text{free}}, \label{effcs}
\end{eqnarray}
where $\sigma_{\text{free}}$ is the free space $NN$ cross section, $\mu$ and $\mu^*$ are the reduced masses of nucleon pairs in free space and medium, respectively. Neglecting the difference between the bare neutron and proton masses, i.e., $m_n=m_p=m_N$, the ratio reduces to
\begin{eqnarray}
    \frac{\mu^*}{\mu}=\frac{1}{m_N}\frac{2m_\tau^*m_{\tau'}^*}{m_\tau^*+m_{\tau'}^*},
    \label{emrde}
\end{eqnarray}
with the nucleon effective mass defined as 
\begin{eqnarray}
    m_\tau^*=\Big( 1+\frac{1}{p}\frac{d U^{\tau}_{\text{MDI}}}{d p} \Big)^{-1}.
    \label{m_ImMDI}
\end{eqnarray}
Here $U^{\tau}_{\text{MDI}}$ denotes the single-particle potential corresponding to the improved isospin- and momentum-dependent interaction (MDI).
In our previous work \cite{2022PhRvC.106f4332H}, we found that the medium effects of the effective mass in the cross section dominated the collective flow of emitted nucleons, suggesting that the adoption of the effective cross section with the effective mass correction is reasonable. 
However, when comparing the influence of the $G$ matrix with that of the effective mass, an opposite trend is observed, indicating that the correction of the effective mass alone is not sufficient. 
This comparison underscores the necessity of incorporating additional medium effects beyond the effective mass correction. Therefore, microscopic many-body calculations of in-medium $NN$ scattering cross sections are crucial to achieving a more accurate theoretical description of heavy-ion collisions (HICs).

\begin{figure*}
    \centering
    \includegraphics[width=0.95\textwidth]{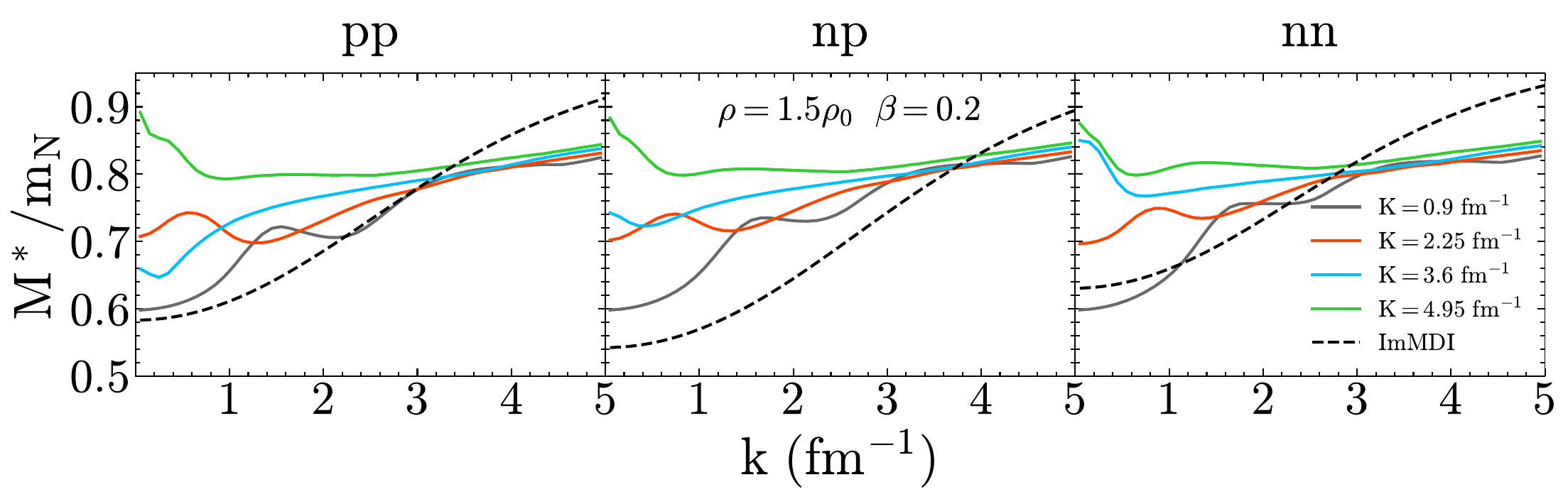}
    \caption{The effective masses of nucleon pairs from Eqs. (\ref{m_ImMDI}) and (\ref{efm}). The left, middle and right panels correspond to the  $pp$, $np$ and $nn$, respectively. The black, red, blue and green lines correspond to the total momenta 0.9, 2.25, 3.6 and 4.95 $\text{fm}^{-1}$, respectively. The black dashed line denotes the results calculated using the BHF method. The black dashed line denotes the results from the improved MDI.}
    \label{mstar}
\end{figure*}

The in-medium $NN$ cross section requires two central ingredients, namely the scattering amplitude and the density of states. 
Within the BHF approximation, the effective $G$ matrix, serving as the scattering amplitude, and the effective mass, associated with the density of states, can be obtained self-consistently.
Using the $G$ matrix, the total $NN$ elastic scattering cross section in asymmetric nuclear matter can be expressed as
\begin{eqnarray}
\sigma(\rho,\beta,k,K)&=&\frac{M^{*2}}{16\pi^2\hbar^4}\sum_{S,J}\sum_{L',L}\left[1-(-1)^{S+L+T}\right]^2\nonumber\\
&\times&\frac{2J+1}{4\pi}\left|G_{L'L}^{SJ}(\rho,\beta,k,K)\right|^2 \ , \label{sig}
\end{eqnarray}
where $G_{L'L}^{SJ}=\langle k, L'SJ|G(e_{2})|k, LSJ\rangle$ denotes the on-shell $G$-matrix element in the partial-wave representation. Here, $T$, $S$, $L$, and $J$ denote the total isospin, total spin, orbital angular momentum and total angular momentum of the two scattering nucleons, respectively. In addition to these quantum numbers, the in-medium cross sections also depend on density and isospin asymmetry, since both the scattering amplitude and the density of states are implicit functions of these quantities. It is also worth emphasizing the role of total momentum $K$, which originates from the exact treatment of the intermediate scattering states \cite{2021PhRvC.103c4316S}. When the total momentum is included, the calculation of the effective mass of the two-nucleon system in medium requires careful consideration \cite{1997PhRvC..55.3006S,1990AnPhy.202...57S}. The effective mass is defined through the dependence of the total energy of the pair, $e_{2}=e^{\tau}_{\emph{\textbf{p}}_{1}}+e^{\tau'}_{\emph{\textbf{p}}_{2}}$ on the relative $k=|\emph{\textbf{p}}_{1}-\emph{\textbf{p}}_{2}|/2$:
\begin{eqnarray}
M^*= \Big(\frac{1}{2k}\frac{d e_{2}}{d k}\Big)
^{-1}\label{efm} \ .
\end{eqnarray}
Where $e^{\tau}_{\textbf{p}_{1}}=\hbar^2\textbf{p}_{1}^2/(2m_N)+U^\tau_{\text{BHF}}$ is the single particle energy in the BHF approach, with $U^\tau_{\text{BHF}}$ denoting the corresponding single-particle potential. This definition is equivalent to Eq.~(\ref{emrde}) only when the momenta of the colliding particles are equal ($|\textbf{p}_{1}|=|\textbf{p}_{2}|$) or when the single-particle potential is parabolic with $U^\tau_{\text{MDI}}$ in Eq.~(\ref{emrde}) replaced by $U^\tau_{\text{BHF}}$. In general, however, the two are not identical, and the definition (\ref{efm}) also contains the influence of the total momentum $K$.
In Fig. \ref{mstar}, we present the nucleon effective mass defined by Eqs. (\ref{m_ImMDI}) and (\ref{efm}) at zero temperature. The solid line represents the results calculated using the BHF approach, while the dashed line corresponds to the result from the improved MDI. From the figure, it can be observed that the BHF results exhibit relatively complex momentum-dependent behavior, but overall, the effective mass increases with the total momentum. In contrast, the improved MDI results gradually exceed those from BHF as the momentum $k$ rises.

\section{RESULTS AND DISCUSSION}

In this section, we focus on investigating the in-medium effects of $NN$ cross sections based on microscopic calculations in heavy-ion collisions. Specifically, we examine their impact on several key observables, including the neutron-to-proton ($n/p$) ratio, the difference between neutron and proton transverse flows, the neutron-proton differential collective flow, the multiplicities of $\pi^-$ and $\pi^+$ mesons, and the resulting $\pi^-/\pi^+$ ratio. 
Particular attention is paid to the medium modifications of the scattering amplitude and the density of states in the cross section, as well as to the influence of accurate treating the total momentum in the BBG equation. 
By examining these observables, we aim to gain valuable insights into the behavior of nuclear matter in HICs.

\begin{figure*}
    \centering
    \includegraphics[width=0.9\textwidth]{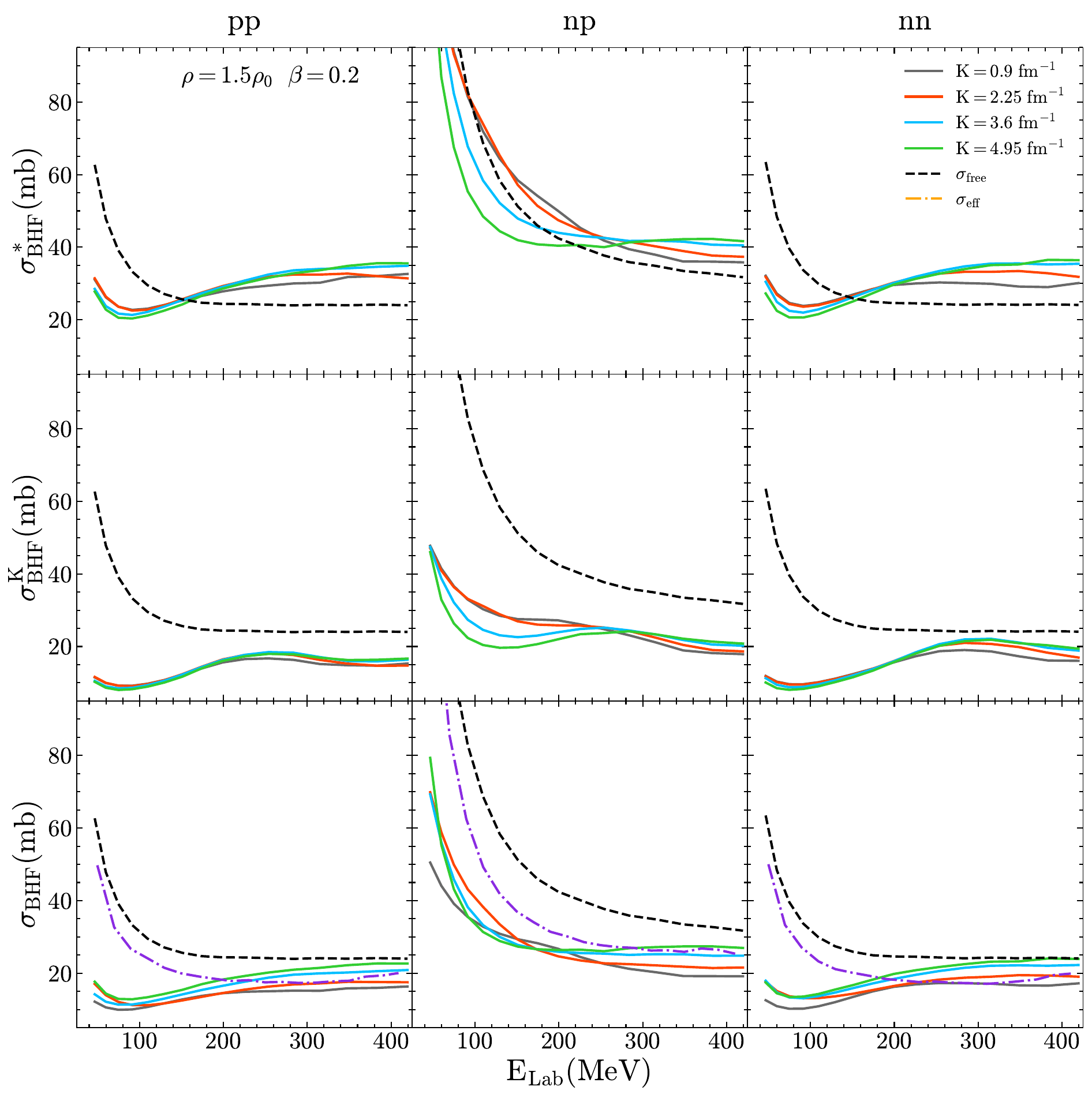}
    \caption{The calculated in-medium $NN$ cross section within BHF approach as a function of the
    incident laboratory energy $E_{\rm Lab}$. The left, middle and right panels correspond to the in-medium $pp$, $np$ and $nn$ cross sections, respectively. While the upper, middle and lower panels show the results with the microscopic effective $NN$ cross section $\sigma_{\rm BHF}^*$, microscopic effective $NN$ cross section $\sigma^K_{\rm BHF}$ and microscopic $NN$ cross section $\sigma_{\rm BHF}$, respectively. The black, red, blue and green lines correspond to the total momenta 0.9, 2.25, 3.6 and 4.95 $\text{fm}^{-1}$, respectively. The black dashed line denotes the free-space $NN$ cross section. The blue-violet dash-dotted line represents the effective cross section $\sigma_{\text{eff}}$, which is output from the IBUU simulations and obtained through statistical averaging using Eq.~(\ref{effcs}), thereby effectively accounting for these temperature-related effects.
    }
    \label{cs}
\end{figure*}

First, the in-medium proton-proton ($pp$), neutron-proton ($np$), and neutron-neutron ($nn$) cross sections obtained based on the BHF approaches are shown in Fig. \ref{cs}. 
These cross sections serve as crucial inputs in our analysis and play a significant role in understanding the behavior of nuclear matter in HICs.
Here, we present the microscopic in-medium $NN$ cross sections calculated at various total momenta $K$, using a representative density of $\rho = 1.5\rho_0$ and isospin asymmetry of $\beta = 0.2$ as an example. The dashed lines represent the $NN$ cross sections in free space. By substituting the effective mass of the collision pairs $M^*$ with the vacuum mass, while retaining the same scattering amplitude $G_{L'L}^{SJ}$ in Eq.~(\ref{sig}), we obtain the approximate cross sections denoted as $\sigma_\text{BHF}^*$, which allow us to isolate the medium effects arising from the scattering amplitude. This approximate $NN$ cross section $\sigma_\text{BHF}^*$ surpasses the free-space value, indicating the medium enhancement of the scattering amplitude \cite{1997PhRvC..55.3006S,2022PhRvC.106f4332H}. A comparison between $\sigma_\text{BHF}$ and $\sigma_\text{BHF}^*$ also displayed in Fig. \ref{cs} elucidates the distinct effects of the scattering amplitude and the density of states, highlighting the significance of the density of states in the cross sections. 
Furthermore, as shown in Fig. \ref{cs}, $\sigma_\text{BHF}^*$ in the $nn$ and $pp$ channels are significantly suppressed at low energies but enhanced at high energies. This behavior mainly originates from the effects of three-body forces (TBF). 
The BHF calculations including a phenomenological TBF in Ref. \cite{baldo} show a similar trend. In addition, our previous work \cite{2022PhRvC.106f4332H} demonstrated that this behavior becomes more pronounced at higher densities, consistent with the increasing importance of TBF effects in this regime. An analysis of the single particle potential $U(k)$ in BHF aproach further shows that the change of $U_{nn/pp}(k)$ with and without TBF is significantly stronger than that of $U_{np/pn}(k)$, which consequently explains why a similar behavior does not appear in the $np$ cross section. 

Moreover, to properly incorporate the medium effects, we additionally illustrate the variation of the cross section with total momentum $K$, with $K$ determined from an exact description of the scattering nucleon states. It is observed that the medium effect becomes weaker as the total momentum $K$ increases in the case of $\sigma_\text{BHF}$. 
To investigate the role of the total momentum $K$ in the cross section, we present the microscopic effective cross section $\sigma^K_{\text{BHF}}$ in Fig. \ref{cs}. Here $\sigma^K_{\text{BHF}}$ is obtained by replacing $M^*/m_N$ with $\mu^*/\mu$ in Eq. (\ref{emrde}), but substituting $U^\tau_{\text{BHF}}$ for $U^\tau_{\text{MDI}}$.  
It should be noted that the definition of $\sigma^K_{\text{BHF}}$ still contain the $K$ dependence from the scattering amplitude. The present approximation $\sigma^K_{\text{BHF}}$ is motivated by two considerations. On the one hand, the $G$ matrix is obtained from a self-consistent calculation, in which the total momentum $K$ of the colliding pairs plays an important role~\cite{2021PhRvC.103c4316S}. Consequently, it is not straightforward to remove the contribution of $K$ solely from the scattering amplitude. More importantly, the medium effects contained in the scattering amplitude are much weaker than those associated with the effective mass. On the other hand, we also aim to investigate the difference between using the effective mass of the colliding pairs obtained from a full calculation (\ref{efm}) and adopting the reduced effective mass approximation (\ref{emrde}). It can be seen that though the $K$ dependence of $\sigma^K_{\text{BHF}}$ is significantly reduced, especially for the $nn$ and $pp$ channels, the resulting cross sections remain only slightly smaller than those of $\sigma_{\text{BHF}}$. 

For comparison, Fig. \ref{cs} also shows the cross section corrected by the $R$-factor from Eq. (\ref{effcs}). In HIC simulations, the momentum distributions of nucleons in the projectile and target nuclei are typically initialized at zero temperature. These zero-temperature initial distributions gradually evolve away from their initial state during the subsequent dynamical evolution. In the IBUU model, the resulting medium effects enter the scattering cross section through the factor $\mu^*/\mu$. However, these effects cannot be simply described in terms of a definite temperature. Although BHF calculations can, in principle, include temperature effects \cite{2020PhRvC.101f5801S}, the temperature is treated as a well-defined thermodynamic quantity within the BHF framework. During the nonequilibrium transport evolution, however, the system lacks a globally well-defined temperature. Consequently, self-consistently incorporating the effective temperature effects present in the IBUU model into the temperature-dependent cross section $\sigma_{\text{BHF}}(T)$ remains challenging. Therefore, as a first approximation in the present work, we adopt the cross section $\sigma_{\text{BHF}}$ calculated in the zero-temperature limit and leave the investigation of effective temperature effects on the in-medium cross sections for future work. In contrast, the cross section $\sigma_{\text{eff}}$ shown in Fig.~\ref{cs} is extracted from the transport simulations and effectively accounts for these temperature-related effects.

In the present work, the in-medium cross section obtained within the BHF approach depends on both the total momentum $K$ of the  colliding nucleon pair and the relative momentum $k$ (corresponding to the laboratory energy $E_{\text{Lab}}=2\hbar^2k^2/m_N$), and implicitly on the density $\rho$ and isospin asymmetry $\beta$ of the medium. The latter three quantities are already included in the IBUU model. To incorporate the $K$ dependence in the IBUU framework, once a colliding nucleon pair with momenta $\textbf{p}_1$ and $\textbf{p}_2$ is identified in a medium with given density $\rho$ and isospin asymmetry $\beta$, the total and relative momenta are defined as $K=|\textbf{p}_1+\textbf{p}_2|$ and $k=|\textbf{p}_1-\textbf{p}_2|/2$, respectively. The corresponding scattering cross section $\sigma(\rho,\beta,k,K)$ in Eq. (\ref{sig}) obtained from BHF approach, is then introduced in the IBUU transport model. It should be noted that, in the practical implementation within the IBUU model, the cross section $\sigma_{\text{BHF}}$ is taken from the parameterization given in our previous work \cite{2022PhRvC.106f4332H}, whereas the cross sections $\sigma_{\text{BHF}}^*$ and $\sigma_{\text{BHF}}^K$ are implemented in the IBUU model via interpolation. As shown Ref. \cite{2022PhRvC.106f4332H}, this parameterization leads to no noticeable difference in heavy-ion collision observables compared with directly interpolating the calculated $\sigma_{\text{BHF}}$.

\subsection{Free nucleons distribution and nuclear stopping}

\begin{figure}
    \centering   
    \includegraphics[width=0.45\textwidth]{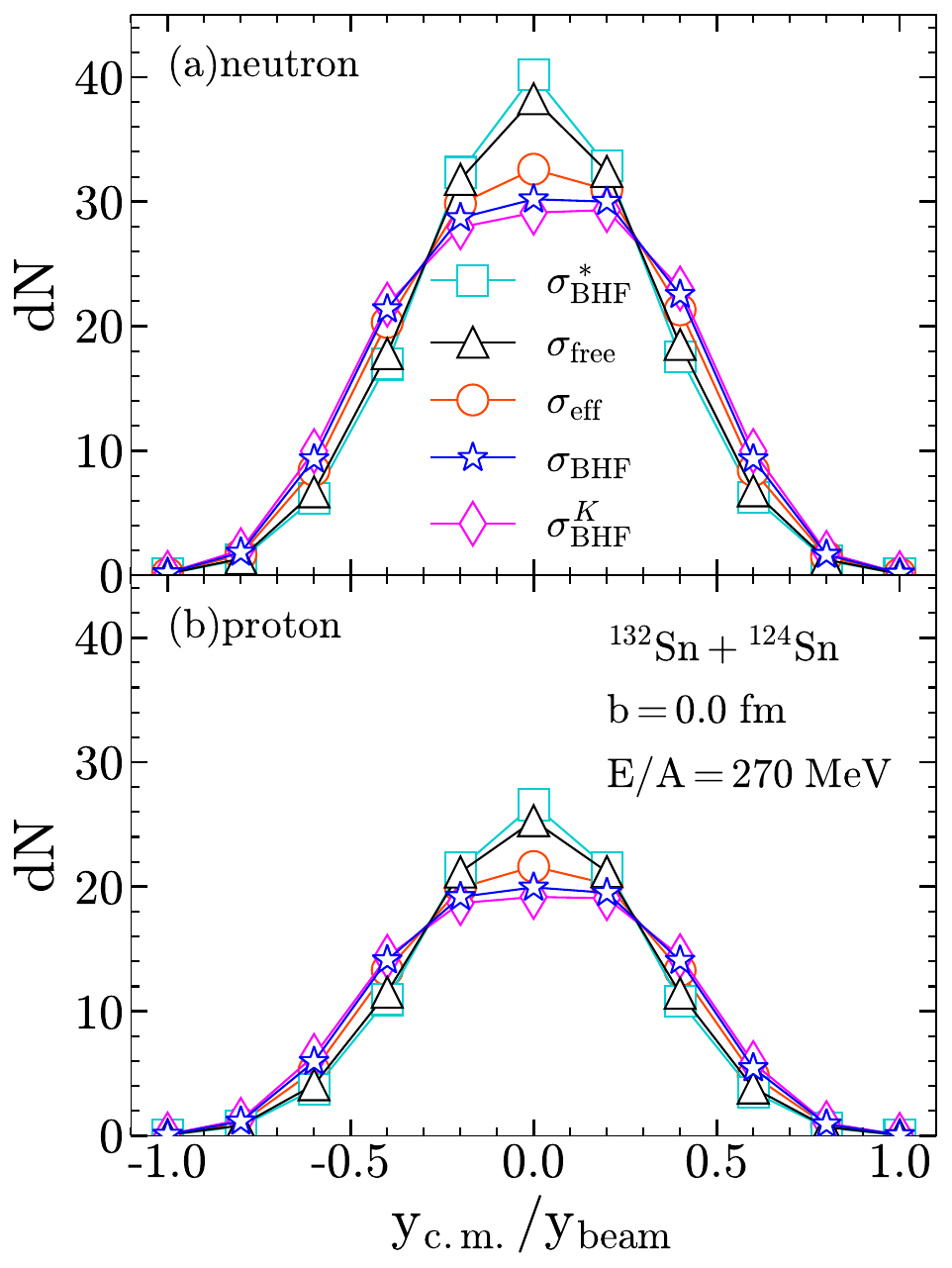}
    \caption{Rapidity distributions of neutrons (a) and protons (b) in $^{132}\text{Sn}+^{124}\text{Sn}$ reaction at 270 MeV/nucleon, which are simulated with the free-space cross section $\sigma_{\text{free}}$, effective cross section $\sigma_{\text{eff}}$, microscopic cross section $\sigma_{\text{BHF}}$ and microscopically effective cross section $\sigma^*_{\text{BHF}}$ and $\sigma^K_{\text{BHF}}$.}
    \label{dN_np}
\end{figure}

\begin{figure}
    \centering   
    \includegraphics[width=0.45\textwidth]{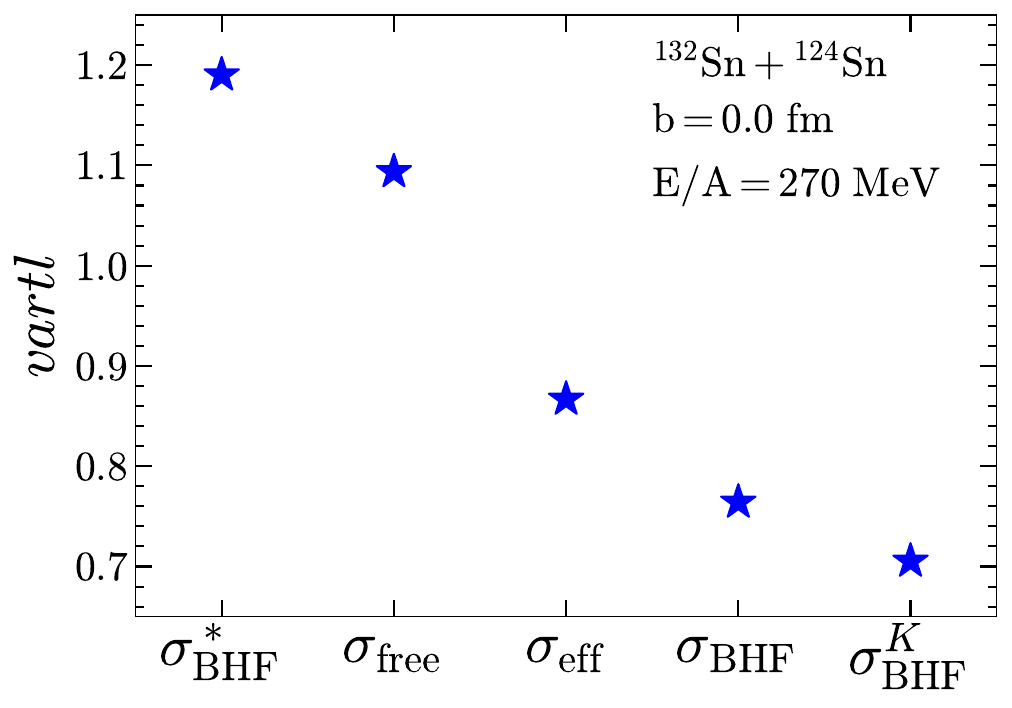}
    \caption{The proton $vartl$ for central $^{132}\text{Sn}+^{124}\text{Sn}$ at 270 MeV/nucleon for the impact parameter $b=0.0\ \text{fm}$ as a function of the cross section.}
    \label{stopping}
\end{figure}
 
Before analyzing the $n/p$ ratio and collective flows, we first investigate in Fig.~\ref{dN_np} the influence of in-medium cross sections on the rapidity distributions of final-state free neutrons and protons. 
We identify free nucleons as those with local densities less than in $\rho_0/8$ the final state of the reaction. 
These results are obtained from transport simulations of the $^{132}\text{Sn}+^{124}\text{Sn}$ reaction at an incident beam energy of 270 MeV/nucleon and an impact parameter of $b = 0.0\ \text{fm}$. 

In the present work, five types of scattering cross sections are considered: $\sigma_{\text{free}}$, the free-space $NN$ cross section; $\sigma_{\text{eff}}$, the effective cross section corrected by the $R$-factor from Eq.~(\ref{effcs}); $\sigma_{\text{BHF}}$, the microscopic cross section derived from the BHF approach; $\sigma_{\text{BHF}}^*$, an approximate cross section calculated by replacing $M^*$ with vacuum mass in Eq.~(\ref{sig}); and $\sigma^K_\text{BHF}$, which neglects the total momentum $K$ dependence of the effective mass of the two-nucleon system, i.e., the expression in Eq.~(\ref{efm}) is approximated by the reduced effective mass of the nucleon pairs. This last case is introduced to isolate the contribution of the total momentum $K$, via the density of states, to observables in HICs.

Larger $NN$ cross sections increase the collision rate, leading to more free nucleons emitted at mid-rapidity.
Due to the enhancement of the cross section by the in-medium $G$-matrix, more free nucleons are produced when using cross section $\sigma^*_{\text{BHF}}$ compared to the case with the free-space cross section $\sigma_{\text{free}}$. 
In contrast, the cross sections $\sigma_{\text{eff}}$ and $\sigma_{\text{BHF}}$, which include effective mass effects, suppress the scattering probability and result in fewer emitted nucleons. As the cross section decreases, the longitudinal rapidity distribution of the emitted nucleons becomes broader, which signifies a reduction in nuclear stopping. 
The enhancement in nucleon yield from $\sigma^*_{\text{BHF}}$ is also noticeably smaller than the suppression caused by $\sigma_{\text{BHF}}$, indicating that the medium effect on the $G$-matrix might be negligible to some extent. 
Nevertheless, for a more accurate and detailed study, it is still necessary to take into account the medium modification of the scattering matrix. 
Given the dominant role of the effective mass in the suppression of in-medium cross-section, we also consider a case where the effective mass is assumed to be independent of the total momentum $K$. In this approximation, relative to the $\sigma_{\text{BHF}}$ results, the influence of the total momentum $K$ on the emitted nucleons manifests medium effects of comparable magnitude but opposite tendency to those embedded in the scattering amplitude (the result of $\sigma^*_{\text{BHF}}$), as shown in Fig.~\ref{dN_np}. For the rapidity distributions, simulations based on $\sigma_{\text{BHF}}$ produce fewer free nucleons than those obtained with the empirical effective cross section $\sigma_{\text{eff}}$. The discrepancy remains modest, thereby confirming the reliability of $\sigma_{\text{eff}}$.

To further investigate the in-medium effects of microscopic scattering cross sections $\sigma_{\text{BHF}}$ in HICs, we analyze the nuclear stopping power, quantified by the observable $vartl$~\cite{2023PhRvC.108f4603S,2006EPJA...30...31A,2018NuScT..29..177L,2024PhLB..85338661T}. 
This quantity characterizes the extent to which the energy and momentum of the projectile nuclear are dissipated and redistributed during the collision process.
The $vartl$ is defined as the ratio of the variance of the transverse rapidity distribution to that of the longitudinal one~\cite{2006EPJA...30...31A,2004PhRvL..92w2301R}:
\begin{equation}
vartl = \frac{\langle y_x^2\rangle}{\langle y_z^2\rangle} \ .
\end{equation}
The transverse and longitudinal rapidities, $y_x$ and $y_z$, are given by:
\begin{equation}
y_{x} = \frac{1}{2} \ln\frac{E + p_{x}}{E - p_{x}},\ y_{z} = \frac{1}{2} \ln\frac{E + p_{z}}{E - p_{z}} \ ,
\end{equation}
where $E$ is the total energy of the nucleon and $p_{x(z)}$ is the momentum component along the transverse ($x$) or longitudinal ($z$) direction. When $vartl \approx 1.0$, it indicates that the projectile is completely stopped by the target. 
The value of $vartl$ is strongly affected by the in-medium $NN$ scattering cross section, making it a sensitive probe of medium modifications to the cross section. 
Figure~\ref{stopping} shows the calculated $vartl$ values for protons. 
The horizontal axis corresponds to the five types of elastic $NN$ cross sections labeled $\sigma^*_{\text{BHF}}$, $\sigma_{\text{free}}$, $\sigma_{\text{eff}}$, $\sigma_{\text{BHF}}$ and $\sigma^K_{\text{BHF}}$. 
A stronger medium effect suppresses the $NN$ cross section, leading to a corresponding decrease in the $vartl$. 
For $\sigma^*_{\text{BHF}}$, the enhancement of the scattering matrix in the medium results in a cross section larger than that in free space, yielding $vartl > 1$, i.e., an over-stopping of protons. 
In contrast, for $\sigma_{\text{BHF}}$ and $\sigma^K_{\text{BHF}}$, the reduction of the cross section caused by the effective mass leads to $vartl < 1$. Moreover, a clear difference can be observed between $\sigma_{\text{BHF}}$ and $\sigma^K_{\text{BHF}}$, indicating that the contribution of the total momentum $K$ through the density of states to the nuclear stopping is comparable to the medium effect arising from the scattering amplitude (see also the discrepancy between $\sigma^*_{\text{BHF}}$ and $\sigma_{\text{free}}$). Therefore, the nuclear stopping may serve as a useful observable to constrain the in-medium $NN$ cross sections.

\subsection{Neutron and proton collective flows}

\begin{figure}
    \centering   
    \includegraphics[width=0.45\textwidth]{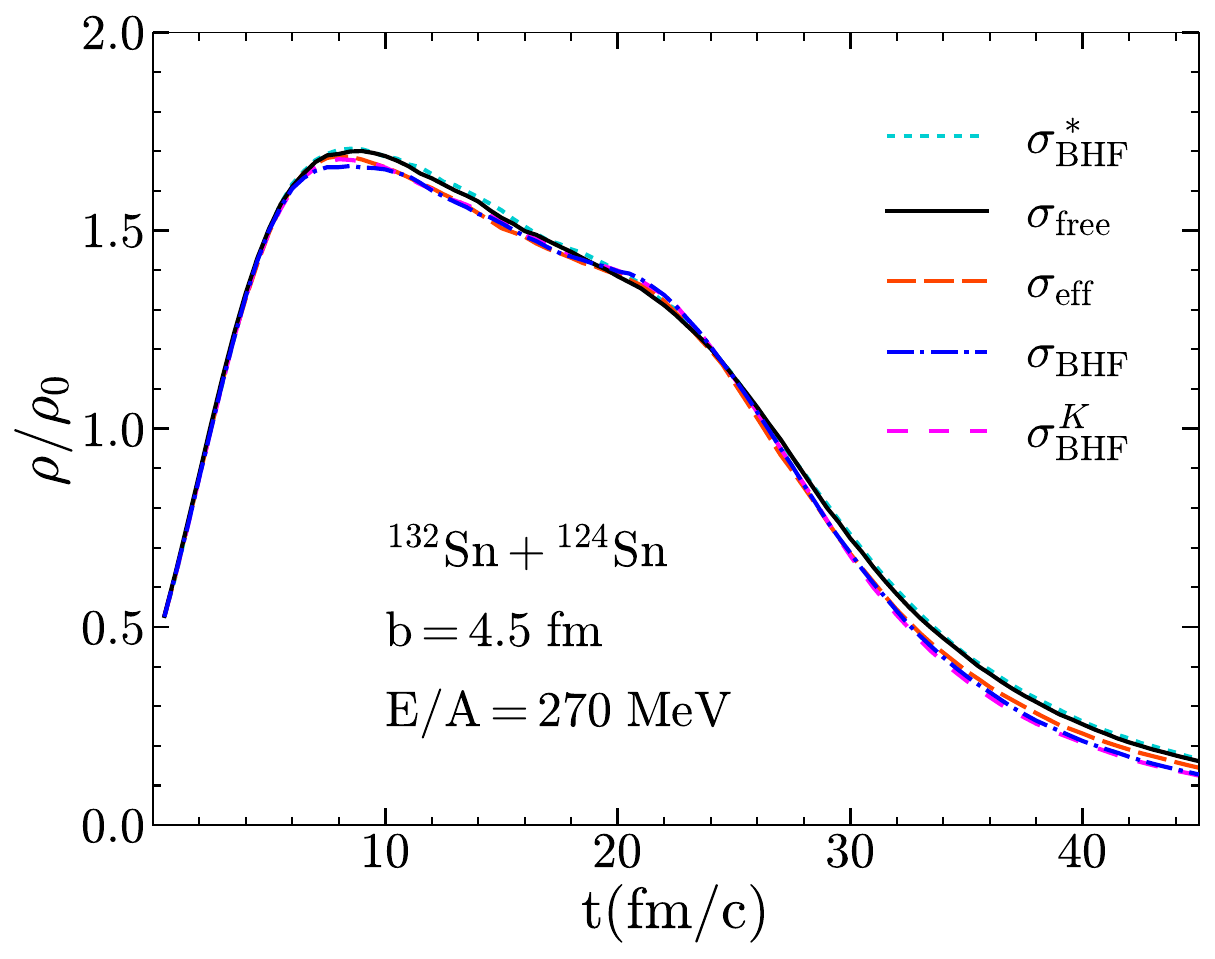}
    \caption{Time evolution of central density in semi-central reaction of $^{132}$Sn+$^{124}$Sn at 270 MeV/nucleon, which are simulated with the free-space cross section $\sigma_{\text{free}}$, effective cross section $\sigma_{\text{eff}}$, microscopic cross section $\sigma_{\text{BHF}}$ and microscopically effective cross section $\sigma^*_{\text{BHF}}$ and $\sigma^K_{\text{BHF}}$.}
    \label{rho_t}
\end{figure}

First, Fig. \ref{rho_t}, shows the time evolution of the central nucleon density in semi-central collisions of $^{132}$Sn+$^{124}$Sn at 270 MeV/nucleon for the five different cross-section scenarios. The maximum central density increases with increasing cross section. A larger cross section enhances the friction between the colliding nuclei, slowing their interpenetration and prolonging the compression stage, which results in the production of more free nucleons.

\begin{figure*}
    \centering   
    \includegraphics[width=0.9\textwidth]{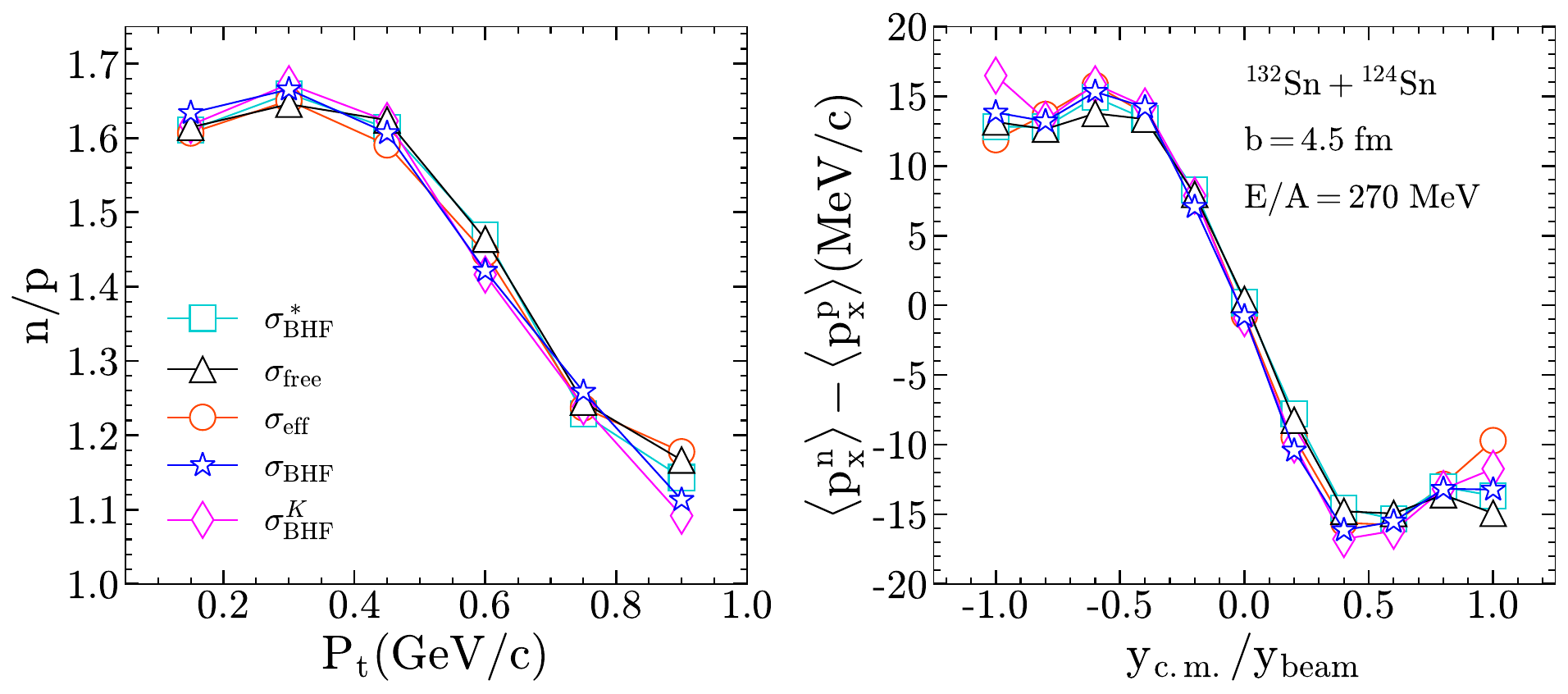}
    \caption{The $n/p$ ratio (left panel) and the difference of the neutron-proton transverse flow (right panel) in semi-central reaction of $^{132}$Sn+$^{124}$Sn at 270 MeV/nucleon. Left panel: The $n/p$ ratio as a function of the transverse momentum. Right panel: The difference of the neutron-proton transverse flow as a function of rapidity.}
    \label{np-px}
\end{figure*}

The $n/p$ ratio and the difference in the neutron-proton transverse flow are highly sensitive probes for investigating the high-density behavior of the symmetry energy \cite{2014PhRvC..90d4605G,2022PhRvC.106a4604H,2011PhLB..700..139C,2006EPJA...30..153D}. 
Moreover, the emitted nucleon multiplicity and nucleon transverse flow are also strongly affected by the in-medium $NN$ elastic scattering cross section \cite{2022PhLB..82837019L,2007PhRvC..75c4615Z}. 
Accordingly, examining the influence of in-medium $NN$ cross sections on the $n/p$ ratio and neutron-proton transverse flow difference plays a pivotal role in constraining the symmetry energy of dense nuclear matter.
Figure \ref{np-px} displays the transverse-momentum and the rapidity distribution of the $n/p$ ratio for free nucleons and the rapidity distribution of the neutron-proton transverse flow difference.  
As seen From Fig. \ref{np-px}, both the $n/p$ ratio and the neutron-proton transverse flow difference are weakly sensitive to the $NN$ elastic cross section. 
Previous studies ~\cite{2018PhRvC..97c4604F,2007PhLB..650..344Y} have shown that the neutron-proton ratio $n/p$ and the difference $\langle p_x^n\rangle-\langle p_x^p\rangle$ between neutron and proton transverse flow exhibit a strong dependence on the symmetry energy. Therefore, these observables are regarded as cleaner probes for investigating the symmetry energy. Their insensitivity to the in-medium $NN$ cross section makes them more reliable and robust quantities for studying the symmetry energy behavior in HICs.

\begin{figure*}
    \centering    
    \includegraphics[width=0.9\textwidth]{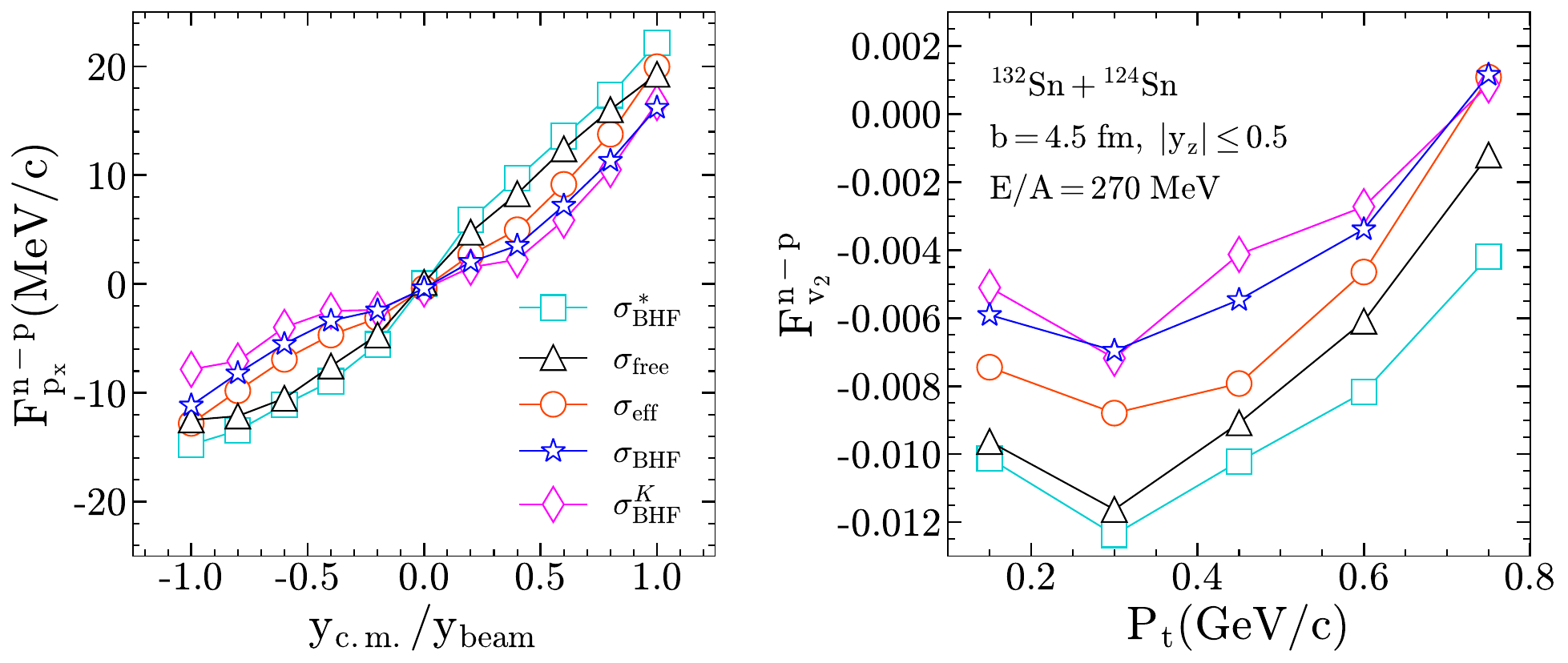}
    \caption{Same reaction as Fig. \ref{np-px}, but for the results of the neutron-proton differential transverse (left panel) and elliptic (right panel) flow.}
    \label{dflow1}
\end{figure*}

The neutron-proton differential collective flows are also highly sensitive probes for studying the high-density behavior of symmetry energies, as they are closely related to isospin fractionation effects \cite{2022PhRvC.106a4604H,2009PrPNP..62..419C}. 
These observables highlight the isospin-dependent dynamics of neutrons and protons in HICs. 
The neutron-proton differential transverse and elliptic flow are defined as \cite{2000PhRvL..85.4221L,2006PhRvC..74f4617Y,2014PhLB..735..250X}
\begin{eqnarray}
    F^{n-p}_{p_x}&=&\frac{1}{N}\sum_{i=1}^{N}p_{ix}\omega_{i}\nonumber\\
    &=&\frac{N_n}{N}\langle p^{n}_{x}\rangle-\frac{N_p}{N}\langle p^{p}_{x}\rangle \ , \label{dtf}
\end{eqnarray}
and
\begin{eqnarray}
    F^{n-p}_{v_2}&=&\frac{1}{N}\sum_{i=1}^{N}\frac{p_{ix}^2-p_{iy}^2}{p_{t}^2}\omega_i\nonumber\\
    &=&\frac{N_n}{N}v_{2}^n-\frac{N_p}{N}v_{2}^p \ , \label{dvf}
\end{eqnarray}
where $N$, $N_n$ and $N_p$ denote the numbers of free nucleons, neutrons, and protons, respectively, $p_{ix}$ is the transverse momentum of the free nucleon, $\langle p^{n(p)}_{x}\rangle$ and $v_{2}^{n(p)}$ are the transverse and elliptic flow of neutrons (protons); and the weight factor $\omega_i=1(-1)$ for the neutrons (proton). 
From Eqs.~(\ref{dtf}) and (\ref{dvf}), it is evident that the neutron–proton differential flow contains valuable information not only on the difference between neutron and proton collective flows but also on the n/p ratio of free nucleons. Figure~\ref{dflow1} presents the neutron–proton differential collective flow, which serves as a sensitive probe of isospin-dependent dynamics in HICs. As shown in the left panel of Fig.~\ref{dflow1}, the differential transverse flow exhibits a clear asymmetry, reflected in the rapidity distributions of $N_n(y)/N(y)$ and $N_p(y)/N(y)$. 
Furthermore, comparison of both panels reveals that the neutron–proton differential flow demonstrates strong sensitivity to the in-medium $NN$ scattering cross sections, even though the $n/p$ ratio and the neutron–proton transverse flow difference remain nearly unaffected by these cross sections.
This behavior can be attributed to more frequent $NN$ collisions associated with a larger cross section, which produce more free nucleons, especially for neutrons (see Fig. \ref{dN_np}). As a consequence, the difference between the numbers of neutrons and protons increases, leading to a stronger $np$ differential collective flow.
In particular, relative to the free-space cross section $\sigma_{\text{free}}$, the microscopic cross section $\sigma_{\text{BHF}}$ induces a more pronounced modification of the differential flow than the commonly used effective cross section $\sigma_{\text{eff}}$. 

Since medium effects in microscopic in-medium $NN$ cross sections originate primarily from modifications to the scattering amplitude and the density of states, we further investigate how these two components individually influence the neutron-proton differential flow. 
For comparison, Fig.~\ref{dflow1} also includes the results obtained with the microscopic effective cross section $\sigma_{\text{BHF}}^*$.  
It is evident from Fig.~\ref{dflow1} that the differential flow calculated with $\sigma_{\text{BHF}}^*$ is closer to that obtained using the free-space cross section $\sigma_{\text{free}}$ than to the result from the complete microscopic cross section $\sigma_{\text{BHF}}$. 
Moreover, the use of $\sigma_{\text{BHF}}^*$ leads to an effect opposite to that produced by $\sigma_{\text{BHF}}$. 
This finding suggests that the medium effect associated with the effective-mass reduction plays a dominant role in shaping the differential flow, while the contribution of the $G$-matrix (scattering amplitude) is weaker and acts in the opposite direction. 
As shown in Fig.~\ref{cs}, the microscopic cross section exhibits a strong dependence on the total momentum $K$ of the two colliding nucleons. 
By comparing the results obtained with $\sigma_{\text{BHF}}$ and $\sigma^K_{\text{BHF}}$, it is evident that the $K$-dependence of the density of states also has a noticeable impact on the neutron-proton differential collective flow.
In the above, we have analyzed in detail how the medium effects in the microscopic $NN$ cross sections influence the neutron-proton differential flow. 
It should be emphasized, however, that flow observables are also sensitive to other factors, such as the equation of state and in-medium inelastic $NN$ scattering. 
Therefore, these observables should be interpreted with caution when used to constrain the in-medium $NN$ elastic cross section.

\subsection{$\pi$ meson emissions}

\begin{figure}
    \centering
    \includegraphics[width=0.45\textwidth]{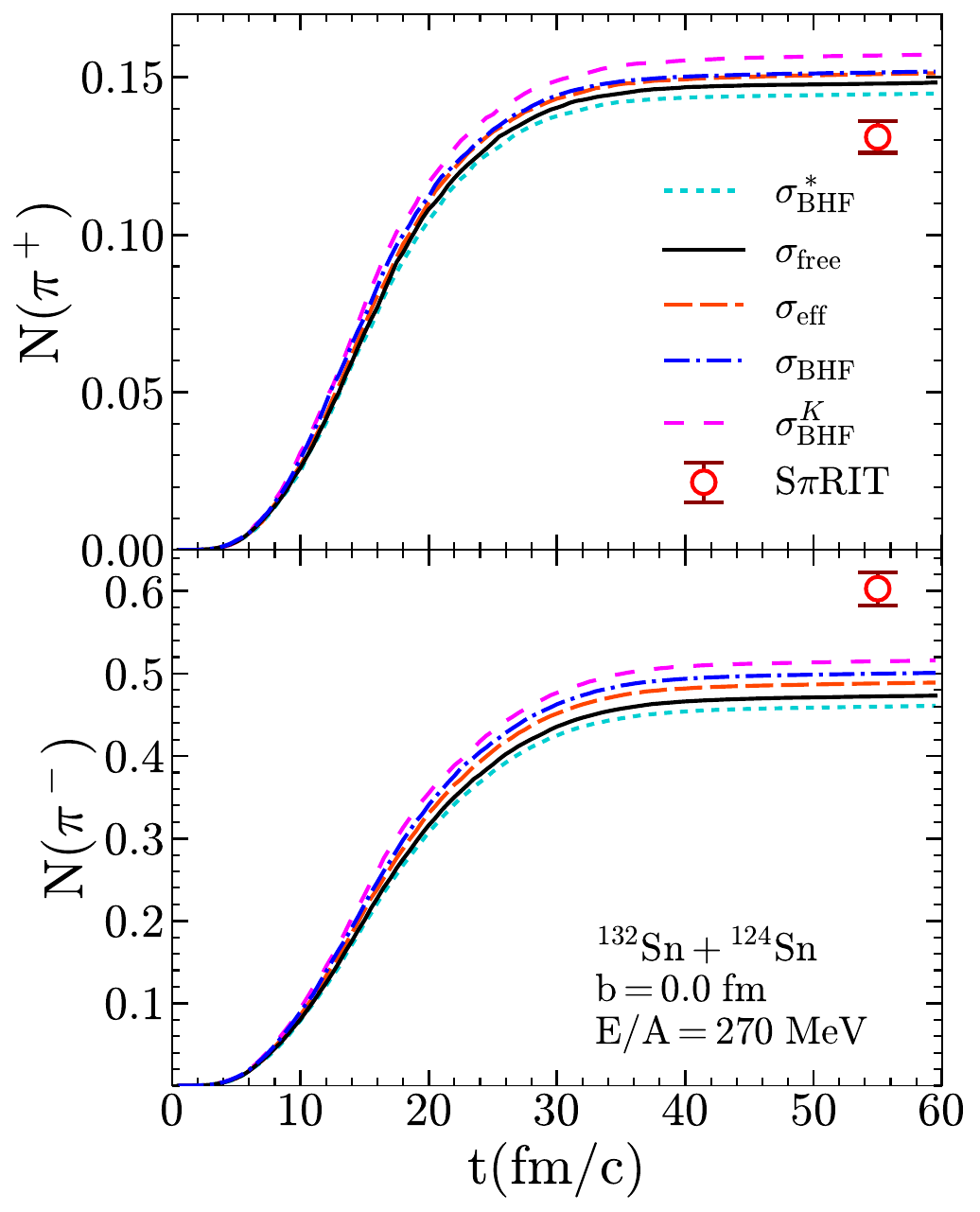}
    \caption{The time evolution of $\pi^-$ and $\pi^+$ multiplicities in the central reaction of $^{132}$Sn+$^{124}$Sn at a beam energy of 270 MeV/nucleon.}
    \label{Npi}
\end{figure}

\begin{figure}
    \centering
    \includegraphics[width=0.45\textwidth]{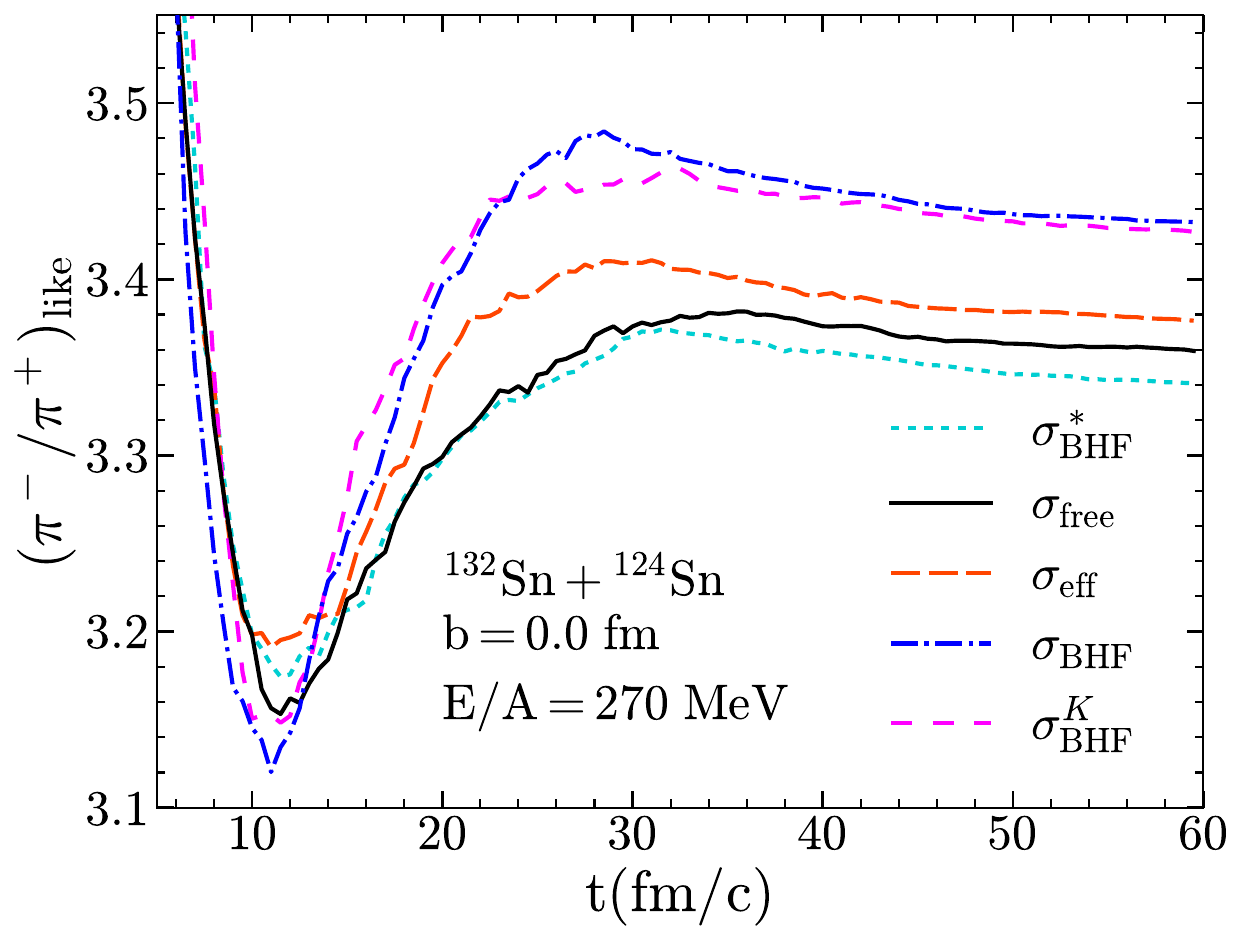}
    \caption{Same as in Fig. \ref{Npi}, but for the time evolution of $(\pi^-/\pi^+)_{\text{like}}$.}
    \label{Rpi}
\end{figure}

The $\pi^-/\pi^+$ ratio has also been recognized as a sensitive probe of the high-density behavior of the symmetry energy \cite{2005PhRvC..71a4608L,2006PhRvC..73c4603Y,2009PhRvL.102f2502X,2002PhRvL..88s2701L,2005PhRvC..72c4613L,2013PhRvC..87f7601X}. 
Since the in-medium $NN$ elastic cross sections influence the pion production process \cite{2010EPJA...46..399Y,2014PhRvC..90d4605G}, the present study focuses on investigating the effects of microscopic in-medium modifications on the final-state pion yields in HICs.

Before examining the $\pi^-/\pi^+$ ratio, we first investigate the influence of the microscopic elastic cross section $\sigma_{\text{BHF}}$ on the $\pi^-$ and $\pi^+$ yields. 
Figure~\ref{Npi} presents the time evolution of the $\pi^-$ and $\pi^+$ multiplicities for a central $^{132}$Sn+$^{124}$Sn collision at $b = 0.0\ \text{fm}$, corresponding to the same reaction system as in Fig.~\ref{np-px}. 
As shown in Fig. \ref{stopping}, a smaller $NN$ elastic cross section reduces nuclear stopping and results in more energetic $NN$ collisions in the fireball matter. In the competition between elastic and inelastic scattering processes, this increases the probability of inelastic scattering and consequently enhances pion production. Consequently, the use of the microscopic cross section $\sigma_{\text{BHF}}$ leads to increased production of both $\pi^+$ and $\pi^-$ compared to the results obtained using the free-space cross section $\sigma_{\text{free}}$ or the effective cross section $\sigma_{\text{eff}}$.
In contrast, the microscopically effective cross section $\sigma_{\text{BHF}}^*$ reduces $\pi^+$ and $\pi^-$ yields relative to $\sigma_{\text{free}}$, because $\sigma_{\text{BHF}}^*$, although incorporating medium-modified scattering amplitudes, exceeds $\sigma_{\text{free}}$ above the pion production threshold (see Fig.~\ref{cs}).
In addition, the results indicate that, in general, the in-medium $NN$ cross sections exert a more pronounced effect on the $\pi^-$ yield than on the $\pi^+$ yield, primarily due to the repulsive Coulomb interaction among protons~\cite{2021PhRvC.104d4607G}. However, for $\sigma^K_{\rm{BHF}}$, the difference in its influence on the $\pi^+$ and $\pi^-$ yields is relatively small.
In addition, Fig. \ref{Npi} compares our results with the S$\pi$RIT measurements for $\pi^+$ and $\pi^-$ data~\cite{2021PhLB..81336016J}. The calculated $\pi^+$ yields are higher than the data, while the $\pi^-$ yields are lower. This discrepancy may originate from the strong sensitivity of pion multiplicities to the symmetry energy~\cite{2005PhRvC..71a4608L,2017PhRvC..96d4605Y}, the medium modification of the inelastic cross section~\cite{2010EPJA...46..399Y,2021PhRvC.103c4615Z}, and the strength of short-range correlations~\cite{2022PhRvC.105a1601Y,yang}, none of which are adjusted in the present work. This comparison highlights the important role of the equation of state and in-medium $NN$ cross sections in HIC simulations and suggests the need for a unified and self-consistent treatment of both.

Figure~\ref{Rpi} shows the time evolution of $(\pi^-/\pi^+)_{\text{like}}$ as influenced by different elastic $NN$ cross sections in the central reaction of $^{132}$Sn+$^{124}$Sn at a beam energy of 270 MeV/nucleon. 
In the dynamics of pion resonance productions and decays, the $(\pi^-/\pi^+)_{\text{like}}$ reads \cite{2010EPJA...46..399Y,2005PhRvC..71a4608L}
\begin{eqnarray}
    (\pi^-/\pi^+)_{\text{like}}=\frac{\pi^-+\Delta^-+\frac{1}{3}\Delta^0}{\pi^++\Delta^{++}+\frac{1}{3}\Delta^+} \ . 
\end{eqnarray}
This ratio naturally becomes the final $\pi^-/\pi^+$ ratio after all $\Delta$ resonance have decayed. 
As shown in Fig.~\ref{Rpi}, the $(\pi^-/\pi^+)_{\text{like}}$ ratio becomes increasingly sensitive to the elastic $NN$ scattering cross section after $t = 15\ \rm{fm}/c$. 
Furthermore, the value of $(\pi^-/\pi^+)_{\text{like}}$ obtained using the microscopic cross section $\sigma_{\text{BHF}}$ exhibits a more pronounced medium modification compared to that obtained with the effective cross section $\sigma_{\text{eff}}$. 
This is because the medium correction introduced by $\sigma_{\text{BHF}}$ is significantly larger for $\pi^-$ than for $\pi^+$, as illustrated in Fig.~\ref{Npi}. 
In addition, the value of $(\pi^-/\pi^+)_{\text{like}}$ calculated using the microscopically effective cross section $\sigma_{\text{BHF}}^*$ is smaller than that obtained with the free-space cross section $\sigma_{\text{free}}$. 
These observations confirm that the effects of the in-medium scattering amplitude and the density of states act in opposite directions on the $(\pi^-/\pi^+)_{\text{like}}$ ratio. 
Interestingly, as shown in Fig. \ref{Npi}, although both the $\pi^+$ and $\pi^-$ yields obtained with $\sigma^K_{\text{BHF}}$ exhibit noticeable enhancements compared with those from $\sigma_{\text{BHF}}$, their difference remains relatively small. As a result, the final values of $(\pi^-/\pi^+)_{\text{like}}$
obtained with $\sigma_{\text{BHF}}$ and $\sigma^K_{\text{BHF}}$ are found to be very close.
This indicates that the $(\pi^-/\pi^+)_{\text{like}}$ ratio exhibits sensitivity to the in-medium $NN$ cross section, but it is also influenced by multiple other factors, including mean-field potentials, resonance production mechanisms, and decay dynamics. In the present study, the ratio does not vary monotonically with different treatments of the $NN$ cross section. Therefore, while it provides useful information, it should be used with care when attempting to constrain the in-medium $NN$ cross section.

\section{SUMMARY}
In conclusion, we have systematically investigated the in-medium effects of $NN$ cross sections on various observables of emitted particles in HICs, using microscopic calculations within the Brueckner–Hartree–Fock framework. The study focuses on key observables, including nuclear stopping, the neutron-to-proton ($n/p$) ratio, the neutron–proton transverse flow difference, the neutron–proton differential collective flow, the multiplicities of $\pi^-$ and $\pi^+$ mesons, and the corresponding $(\pi^-/\pi^+)_{\text{like}}$ ratio. Special attention is given to disentangling the roles of medium modifications arising from the scattering amplitude, the density of states, and the total momentum $K$ of colliding nucleons.

Our analysis shows that larger in-medium $NN$ cross sections generally enhance the emission of free nucleons and affect nuclear stopping, with the effective mass playing a dominant role in suppressing the cross section. However, it is insufficient to account for only the medium corrections associated with the reduced nucleon effective mass of colliding pairs which is commonly used in IBUU simulations. While the correction from the density of states dominates the in-medium modification of cross sections in HICs, the total momentum ($K$) dependence and the medium effects from the scattering amplitude should not be neglected in accurate evaluations. Moreover, the $n/p$ ratio and the neutron–proton transverse flow difference are largely insensitive to in-medium modifications, confirming their reliability as probes of the symmetry energy. In contrast, the neutron–proton differential collective flow exhibits pronounced sensitivity to in-medium effects, with contributions from both the scattering amplitude and the $K$-dependent density of states.

For pion observables, the use of the microscopic cross section $\sigma_{\text{BHF}}$ enhances the production of both $\pi^+$ and $\pi^-$, with a stronger effect on $\pi^-$ due to the repulsive Coulomb interaction between protons. Interestingly, although $\sigma^K_{\text{BHF}}$ also increases pion yields, the difference between $\pi^+$ and $\pi^-$ production remains small, resulting in final $(\pi^-/\pi^+)_{\text{like}}$ ratios with $\sigma_{\text{BHF}}$ and $\sigma^K_{\text{BHF}}$ that are very close. These findings verify that the medium effects associated with the scattering amplitude and density of states act in opposite directions, highlighting the complex interplay of nuclear medium modifications in HIC observables.

In general, this study provides a comprehensive understanding of how microscopic in-medium modifications of $NN$ cross sections influence both nucleonic and mesonic observables, providing valuable insight into the medium effects in nuclear matter. In the future, observables that are sensitive to the in-medium $NN$ cross sections-such as nuclear stopping, differential collective flow, and the the corresponding $(\pi^-/\pi^+)_{\text{like}}$ ratio-may also serve as effective probes to further constrain the in-medium modifications of $NN$ cross sections.

\section*{Acknowledgments}
The work is supported by National Natural Science Foundation of China (grant Nos.~12375117,12273028, 12494572, 12275322), the Youth Innovation Promotion Association of Chinese Academy of Sciences (Grant No. Y2021414), National Key R\& D Program
of China No. 2024YFE0109802, CAS Project for Young Scientists in Basic Research YSBR-088.

\bibliographystyle{apsrev4-1}  

\bibliography{cross}

\end{document}